\documentclass[12pt]{iopart}
 \usepackage{iopams}
 \usepackage{epsfig}   
 \usepackage{graphics}
 \usepackage[T1]{fontenc}

\newcommand{\be}{\begin{equation}}
\newcommand{\ee}{\end{equation}}
\newcommand{\bea}{\begin{eqnarray}}
\newcommand{\eea}{\end{eqnarray}}

\newcommand{\av}[1]{\langle{#1}\rangle}

\begin{document}
\title{Ricci focusing, shearing, and the expansion rate in an almost homogeneous Universe}
\author{Krzysztof Bolejko, Pedro G. Ferreira}
\address{Astrophysics, University of Oxford, DWB, Keble Road, Oxford,
OX1 3RH, UK}
\ead{\mailto{krzysztof.bolejko@astro.ox.ac.uk}, \mailto{p.ferreira1@physics.ox.ac.uk}}

\begin{abstract}
The Universe is inhomogeneous, and yet it seems to be  incredibly well-characterised by a homogeneous relativistic model. One of the current challenges is to accurately characterise the properties of such a model. In this paper we explore how inhomogeneities may affect the overall optical properties of the Universe by quantifying how they can  bias the redshift-distance relation in a number of toy models that mimic the real Universe.
The models  that we explore
are statistically homogeneous on large scales. We find that the 
effect of inhomogeneities is of order of a few percent, which 
can be quite important in precise estimation of cosmological parameters. We discuss what lessons can be learned to help us tackle a more realistic inhomogeneous universe.

\end{abstract}

\maketitle
\section{Introduction}

Large scale redshift surveys of galaxies reveal a hierarchy of structure that extends out to hundreds of Megaparsecs, yet a homogeneous cosmological model 
 lies  at the core of our understanding of how the Universe evolves. Such an assumption works incredibly well, as can be seen from current analyses of the Cosmic Microwave Background (CMB). The linearly perturbed  Friedmann--Lema\^itre--Robertson--Walker (FLRW)  model can fit observations of the CMB with exquisite precision \cite{wmap7}.

Nevertheless we cannot avoid the fact that the Universe is inhomogeneous, especially on small scales. Indeed, as a first approximation, space looks empty, with nuggets of mass and energy clustered together to form an intricate cosmic web. 
It is quite possible that the smooth description of space emerges as the average behaviour of a fundamentally inhomogeneous universe (cf. \cite{dcc0,dcc1,dcc2,dcc3}). 
This is an operation that we
are familiar with closer to home, when we describe a gas or liquid through their macroscopic properties instead of through a detailed atomistic description. We would like to be able to do the same for the Universe, i.e. to infer the smooth, large scale (macroscopic) properties of space time from cosmological observations without having to worry about the discreteness of its constituents on small scales. 
 We attempt to reconcile such an observation with a homogeneous universe by saying that our local, granular point of view is merely the small scale limit of the description given by the $\Lambda$-Cold Dark Matter ($\Lambda$CDM) cosmology. 
We then assume that on large scales, $\Lambda$CDM is accurately described by a smooth cosmological fluid which on small scales fragments into particle dark matter and forms halos. Some have expressed strong concerns about the validity of such an approach  (cf. \cite{Ellis2011,FSL2011,AK2011}), especially because it is not clear whether the inhomogeneities on small scales will radically affect the large scale dynamics of the Universe 
 (see 
\cite{buch2000} -- \cite{CH2012},
but see also \cite{IsWa2006,GrWa2011,CRT2012} for counterexamples).

There are a number of interesting and important issues that must be addressed when tackling this problem. In this paper we wish to focus on one: how inhomogeneities can affect the inferred optical properties of the Universe. This problem has been looked at before, 
 throughout the 1960s 
\cite{Sach1961} -- \cite{Kant1969}
and 1970s 
\cite{DyRo1972} -- \cite{Wein1976},
and more recently
\cite{KaVB1995} -- \cite{DVD2012},
with a variety of different assumptions and results. 

In this paper we
 look at how different effects can modify the redshift-distance relation, ($z$, $D_A$). One might expect, {\it a priori}, that studying the effect of inhomogeneities on ($z$, $D_A$) is pointless: in 1976, Weinberg \cite{Wein1976} claimed that the overall 
($z$, $D_A$) for an inhomogeneous Universe should, on the whole, match that of the average homogeneous Universe. Views are somewhat divided on the validity of Weinberg's argument and its generality, with cosmologists tending to accept it wholesale, while some relativists are slightly more sceptical and question its assumptions (cf. \cite{ElBD1998}). In this paper we will be completely agnostic.

Our approach is based entirely on the Sachs equations. Within the framework
of these equations, we study different aspects of the redshift-distance relation.
This is a  complementary procedure to what is usually done\footnote{The most commonly-implemented procedure to account for the effect of inhomogeneities on light propagation is the 
magnification matrix approach. In almost all cases, one further assumes 
that light propagates on unperturbed geodesics, and thus the redshift
(to a source located at some fixed comoving distance) is calculated as in the background homogeneous models,
 while inhomogeneities only affect the magnification/demagnification
(for theoretical analysis see \cite{HoWa1998,VaFW2008,FKWV2012},
or for practical implementation in ray tracing codes see \cite{MrS1,MrS2,KLLF2012}).}.
Our approach will be to use a number of different approximate models: high resolution N-body simulations, halo models, and a variety of swiss cheese models with Lema\^{i}tre-Tolman-Bondi inhomogeneities.

While what we find is not, as yet, definitive, it does shed light on how important the effects of inhomogeneities might be. The paper is structured as follows. In Section \ref{maths} we present the Sachs equations and their various limits. In Section \ref{R} we look at the effect of Ricci focusing, in Section
 \ref{wsigma} we study the impact of shear, and in Section \ref{sofz} we attempt to include local fluctuations in the expansion rate. In each of these sections, we need to consider different models for inhomogeneities. In Section \ref{disc} we discuss our results.

\section{Distance redshift relation}
\label{maths}

The Sachs optical equations describe the evolution of optical
quantities, the expansion of the null bundle
$\theta$ and its shear $\sigma$ \cite{Sach1961}
\begin{eqnarray}
\frac{{\rm d} \theta}{{\rm d} s} + \theta^2 + |\sigma|^2 = 
- \frac{1}{2} R_{\alpha \beta} k^{\alpha} k^{\beta}, \label{thevo} \\
\frac{{\rm d} \sigma}{{\rm d} s} + 2 \theta \sigma = 
C_{\alpha \beta \mu \nu} \epsilon^{*\alpha}
 k^{\beta} \epsilon^{*\mu} k^{\nu}. \label{sgmevo}
\end{eqnarray}
where $R_{\alpha \beta}$ and $C_{\alpha \beta \mu \nu}$
 are respectively the Ricci and Weyl tensors, $k^\mu$ is 
 the null vector  
and  $\epsilon^\mu$ is
 perpendicular to  $k^\mu$ and is confined  to a surface tangent to a wave front.
In the gravitational lensing nomenclature, `shear' normally refers to the image distortion
$\gamma$. These two quantities are related but are not the same, 
and in order to avoid any confusion, we will always refer to $\sigma$ as to the {\em shearing} (for a detailed set of equations that relates these two quantities, see \cite{CEFMUU2011}).

The rate of change of the distance depends on the expansion rate
of the light bundle in such a way that the angular diameter distance, $D_A$, satisfies 
\begin{equation}
\frac{{\rm d} }{{\rm d} s} \ln D_A = \theta.
\end{equation}
Recall that the luminosity distance is given by $D_L\equiv(1+z)^2D_A$ \cite{Eth33,E1971}.
Rewriting equation (\ref{thevo}), we find that
\begin{equation}
\frac{{\rm d^2} D_A}{{\rm d} s^2} = - ( |\sigma|^2 + \frac{1}{2} R_{\alpha
\beta} k^{\alpha} k^{\beta}) D_A. \label{dsr}
\end{equation}
We need a set of initial conditions to solve for ($z$, $D_A$),
but more importantly we also need:
(1) the Ricci curvature $R_{\alpha\beta}$; 
(2) the Weyl curvature $C_{\alpha\beta\mu\nu}$;
(3) the transformation from the affine parameter $s$ to the redshift $z$.

Using the Einstein equations 
 ($R_{\alpha \beta} - R g_{\alpha \beta}/2 = \kappa T_{\alpha \beta} + \Lambda g_{\alpha \beta}$)
 (where $\kappa = 8 \pi G$ and $G$ is the gravitational constant),
with the energy momentum tensor of a perfect fluid, one has that
\begin{equation} 
R_{\alpha \beta} k^{\alpha} k^{\beta} = \kappa ( \rho + p) (u_{\alpha} k^\alpha)^2.
\end{equation}
In comoving coordinates, using the definition of redshift $1+z \equiv  (u_{\alpha} k^\alpha)_e/  (u_{\alpha} k^\alpha)_o$
(where the subscripts $e$ and $o$ refer to the instants of emission and observation, respectively),
one obtains
\begin{equation}
R_{\alpha \beta} k^{\alpha} k^{\beta} =
\kappa (\rho + p) (1+z)^2,
\end{equation}
where we have used the freedom of the affine re-parametrization\footnote{The affine parametrization
is conserved with respect to linear transformations, $s \to a s +b$
\cite{PlKr2006}.} to set $(u_{\alpha} k^\alpha)_o = 1$.
For the case of dust (pressureless matter),
the Ricci focusing depends only on matter density along the 
past null cone. From now on, we focus only on this case 
(dark matter can be treated as being pressureless, and for the cosmological constant
we have $\rho_\Lambda + p_\Lambda = 0$), i.e. we assume that the Ricci focusing
is given by
\begin{equation}
R_{\alpha \beta} k^{\alpha} k^{\beta} = \rho  (1+z)^2.
\end{equation}
Unfortunately, there is no `easy' trick for estimating the 
Weyl focusing, and so to calculate this factor one has to solve for the null geodesics.
In fact, solving null geodesics is also important for linking
$s$ with $z$, which can be done by using
the redshift formula together with the null geodesic equations.
In comoving coordinates, we have
\begin{equation}
1+z = (u_{\alpha} k^\alpha)_e/  (u_{\alpha} k^\alpha)_o
 =  k^0_e = \frac{{\rm d} t}{{\rm d} s},
\label{sodz}
\end{equation}
which allows us to link $s$ with $z$. In a homogeneous space time we have
\begin{eqnarray}
&&  \frac{{\rm d} z}{{\rm d} s} = (1+z)^2 H, \nonumber \\
&&  \sigma = 0, \nonumber \\
&&  R_{\alpha \beta} k^{\alpha} k^{\beta} =\rho_0  (1+z)^5,\label{flrwlimit}
\end{eqnarray}
where $H$ is the Hubble rate, and one recovers the textbook redshift-distance relation.
If one assumes that 
 $R_{\alpha \beta} k^{\alpha} k^{\beta} = \alpha \rho_0  (1+z)^5,$
where $\alpha = const$, the Dyer-Roeder formula is recovered \cite{DyRo1972,DyRo1973}.


\section{The role of Ricci focusing}
\label{R}

We will first try to quantify the effect of Ricci focusing. To do so, we need
to solve (\ref{dsr}) in terms of $\rho(z)$,
$\sigma(z)$ and $s(z)$.
In this section, we assume that $\sigma(z)$ and $s(z)$
are exactly the same as in the background FLRW model
[see the first two terms in  (\ref{flrwlimit})].
We wish to study `realistic' density profiles along the line
of sight, and to do so we use density fields from the Millennium simulation \cite{Mill1,Mill2}.

The Millennium simulation is an N-body simulation of the concordance cosmology. It consists of
10,077,696,000 particles of mass $8.6 \times 10^8 M_\odot h^{-1}$
within a cube of volume $(500 h^{-1}$ Mpc$)^3$.
The simulation was performed using the GADGET-2 code \cite{gadget}.
In our calculations, we use the Millennium MField, which is the
dark matter density field put on a $256^3$ grid, 
smoothed with Gaussian kernels \footnote{$\rho = \sum_i m_i W$, where $W \sim \exp[ x^2/\sigma^2]. $}
of size: 1.25 Mpc,
2.5 Mpc, 5 Mpc, and 10 Mpc, labelled $g_{1.25}$, $g_{2.5}$, $g_5$, and $g_{10}$ respectively.
We randomly place an observer in the Millennium box, and then calculate
$10^6$ 
different lines of sight (each in a different direction
from the observer, and for each line of sight we 
randomly select a different observer). 
We assume periodic boundary conditions, so that when the light ray exits the Millennium box, it enters the other side of the box with entry angles the same as the exit angles (to enforce periodic boundary conditions).
 Since the MField consists of density maps on discreet time slices, to get
$\rho$ at any required instant we interpolate between different time slices.

\begin{figure}
\begin{center}
\includegraphics[scale=1.]{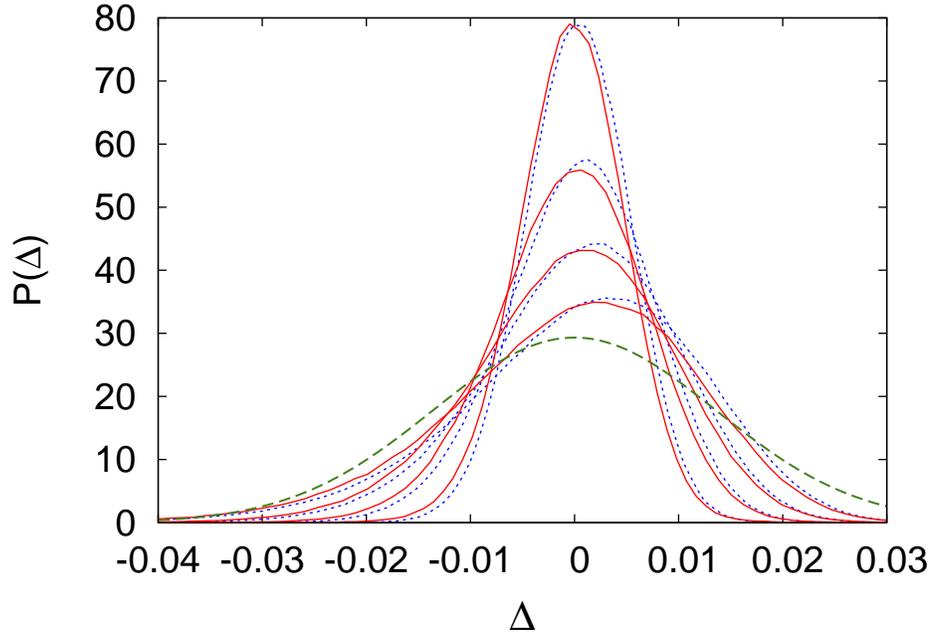}
\caption{PDF of $\Delta$. 
Solid curves (from the top) are the results
obtained by solving Eq. \ref{dsr}
within the Ricci focusing regime,
for the Millennium maps $g_{10}$, $g_{5}$, $g_{2.5}$, and $g_{1.25}$.
The dotted lines are the results 
obtained from (\ref{DWL}) for the same density fields.
The dashed line is the Gaussian PDF with 
the variance calculated from Eq. (\ref {wlsgm}).}
\label{fig1}
\end{center}
\end{figure}

We calculate the light propagation out to $z=1.6$,
solve Eq. (\ref{dsr}) and then write the distance as
a deviation from the expected value in the background model 
(in our case the $\Lambda$CDM model),
\begin{equation}
D_{A}(z)= \bar{D}_{A} ( 1 + \Delta).
\label{dlen}
\end{equation}
Results in terms of the probability distribution function (PDF) of $\Delta$
are presented in Fig. \ref{fig1}.
We find that the larger the smoothing radius, the smaller the variance:
for $g_{10}$ the standard deviation is
 $5.18 \times 10^{-3}$ 
(uppermost solid curve in Fig. \ref{fig1});
for $g_{5}$ the standard deviation is  
$7.28 \times 10^{-3}$
(the solid curve second from the top in Fig. \ref{fig1});
for $g_{2.5}$ the standard deviation is  
$9.52 \times 10^{-3}$ 
(the solid curve third from the top in Fig. \ref{fig1}); and
for $g_{1.25}$ the standard deviation is  
$1.21 \times 10^{-2}$.
It is interesting to compare 
these results with the ones obtained using the weak lensing
formula (in the Born approximation),
\begin{equation}\Delta_{WL} =
- \frac{3 }{2} H_0^2 \Omega_m \int\limits_0^{\chi_e} {\rm d} \chi
\frac{ \chi_e - \chi}{\chi_e} \chi a^{-1} \delta(\chi),
\label{DWL}
\end{equation}
where $\chi$ is the comoving coordinate, $d \chi = dz/H(z)$ and
we use the same  $\delta(z)$ as before.
The results are plotted with dotted lines in Fig. \ref{fig1}.
As shown, the PDFs are very similar to the ones obtained within
the Ricci focusing regime. 
Finally, we calculate the standard deviation 
by squaring the expression above (where the mean is zero by construction)
and replacing $\delta^2$ with the matter power spectrum
\cite{BWM1997},
\begin{equation} \sigma^2_{\Delta} =
\frac{9 }{4} \Omega_m^2 H_0^4 \int\limits_0^{\chi_e} {\rm d} \chi
\left[ (1+z) \frac{ \chi_e - \chi}{\chi_e} \chi \right]^2
\int\limits_0^{\infty} {\rm d} k ~ k \frac{P(k,z)}{2\pi}.
\label{wlsgm}
\end{equation}
In this case, the standard deviation of the distance correction
 is\footnote{Here, in order to compare with the
Millennium simulation, we use the cosmological 
 parameters that were used in the Millennium simulation.
If the WMAP7 set of cosmological parameters is used instead,
then $\sigma_{\Delta} =1.57 \times 10^{-2} $.}
 $1.45 \times 10^{-2}$ (the dashed  curve in Fig. \ref{fig1}
is the Gaussian PDF with mean of zero and standard deviation of $1.45 \times 10^{-2}$).

The above results clearly reveal a pattern. First of all, 
the smaller the smoothing radius, the larger the variance,
although even with the 1.25 Mpc smoothing radius we get
a smaller standard deviation than was obtained within the framework of
linear approximation. Also, note that the maximum of the PDF lies on the demagnification side
 (i.e. $\Delta >0$), so a random object is most likely to be dimmer than on average. Nevertheless, 
in all cases the mean is almost zero:
 $1.20 \times 10^{-4}$,  
 $1.38 \times 10^{-4}$, 
 $1.29 \times 10^{-4}$, 
 $1.44 \times 10^{-4}$,
for $g_{1.25}$,  $g_{2.5}$, $g_{5}$, and $g_{10}$ 
respectively\footnote{Note that, for each different line of sight, we chose
a different observer location.
If we located an observer inside some deep inhomogeneity instead,
such as a void of 
$\delta_{1.25} = -0.84$,
$\delta_{2.5} = -0.78$,
$\delta_{5} = -0.66$,
$\delta_{10} = -0.43$,
then the mean would be 
 $2.54 \times 10^{-4}$,  
 $2.58 \times 10^{-4}$, 
 $2.46 \times 10^{-4}$, 
 $2.07 \times 10^{-4}$.
This suggest that the effect of local inhomogeneity 
on the distance (to $z=1.6$ in our case)
is of order $10^{-4}$.
In this paper we try to avoid the `local' effects
introduced by nearby inhomogeneities,
so that we have a more clear insight to light propagation effects
alone. For papers that study the local bias see
\cite{lvcc}--\cite{lvc5}.}

\begin{figure}
\begin{center}
\includegraphics[scale=1.]{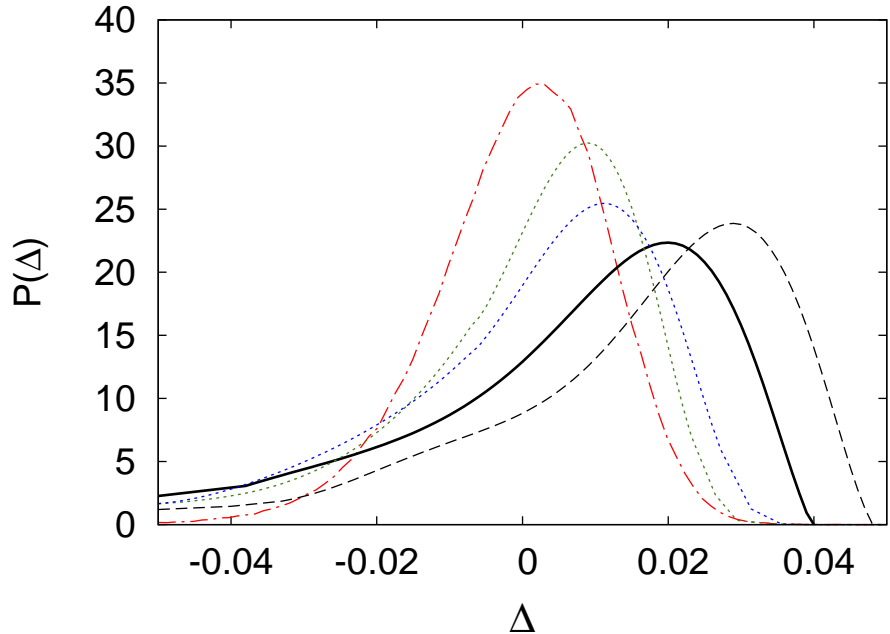}
\caption{The thick solid line is the PDF obtained in the full halo model.
The dashed line is a variation of the halo model, where we switched off the Weyl focusing. The dotted lines (from the top)
are halo models with $\Delta_{SO}$ equal
to $80$ and $40$ respectively (see text for explanation). For comparison, we 
also present the results when the $g_{1.25}$ Millennium map is used 
(dot-dashed line).}
\label{fig2}
\end{center}
\end{figure}

\section{The role of smoothing and Weyl focusing}\label{wsigma}

We now extend our analysis to investigate the role of smoothing and the effect of Weyl focusing on the optical properties of the Universe. 
When dealing with the Weyl focusing, it is not straightforward to link density fields with the Weyl curvature.
An ideal situation would be to have an exact generic solution of the Einstein field equations,
and use that to model the evolution and effects of the Weyl curvature.
Since there is no such a solution, in practise we must always employ some approximations.
Here we will evaluate the Weyl focusing within the {\it halo} model.

Simply put, the halo model considered here consists of a universe filled with
different halos of masses above 
$10^{10} h^{-1} M_\odot$, while the rest of the mass density (in objects of masses below this
threshold) is assumed to be distributed homogeneously.
The halo model we use closely follows the construction 
outlined in detail in Ref. \cite{KKVM2011}.
Each halo is described by the Navarro-Frenk-White (NFW) density profile \cite{NFW},
 
\begin{equation}
\rho(r)= \rho_m(z) \frac{\delta_c}{(r/R)(1+r/R)^2},
\end{equation}
where $R$ is the radius of the halo, defined as
\begin{equation}
R = \left( \frac{M}{(4/3) \pi \rho_m(z) \Delta_{SO}} \right)^{1/3},
\label{halorad}
\end{equation}
with $\rho_m(z) = \rho_0 (1+z)^3$, and $\Delta_{SO} = 180$
(the halo is defined as an overdensity of mass $M$,
whose constant density contrast is 180 with respect to the mean matter density,
cf. \cite{Jenk2001,KKVM2011}).
Integrating $\rho(r)$, and using the expression above, one finds that
\begin{equation} \delta_c = \frac{\Delta_{SO}}{3} \frac{c^3}{\ln(1+c) - c/(1+c)}, \end{equation}
where $c$ is the concentration parameter, whose evolution given in 
\cite{Duff2008} as
\begin{eqnarray}
c = 10.14  [ M /(2 \times 10^{12} h^{-1} M_\odot)]^{-0.081} (1+z)^{-1.01}. 
\end{eqnarray}
The number density of halos at a given instant in time is given by the halo mass function
\[ {\rm d} n = \frac{\rho_m(z)}{M} f {\rm d} \sigma^{-1}, \]
with $f$ and $\sigma$ given by \cite{Jenk2001}
\begin{eqnarray}
&& f(\sigma,z) = 0.301 \exp [ - | \ln \sigma^{-1} +0.67|^{3.82}| ], \nonumber \\
&& \sigma (M,z) = \frac{{\cal G}}{2\pi^2} \int\limits_0^\infty k^2 P(k) W^2(k,M) {\rm d} k, 
\end{eqnarray}
where ${\cal G}(z)$ is the growth factor (this is usually written as $D$, but we use 
$ {\cal G}$ here in order to avoid confusion with distance).

The halo model allows us to  address the problem of smoothing
 --- instead of dealing with a continuous density field,
we now have a discrete set of dark matter halos --- 
although one should keep in mind that, given the characteristics described above, this
is not really a completely `discrete' model.
 It also  allows us to take into account the effect of the
shearing, by assuming that
the Weyl focusing for a particular halo, $i$, is calculated using  
the  Lema\^itre--Tolman (LT) model \cite{Lema1933,Tolm1934} (an exact general 
relativistic model of an spherically symmetric, inhomogeneous, non-stationary
space time) \cite{BrTT2007,BrTT2008}
\begin{equation}
 \mathcal{C}_i\simeq 
C_{\alpha \beta \mu \nu} \epsilon^{*\alpha} k^{\beta} \epsilon^{*\mu} k^{\nu} = \frac{1}{2} \frac{b^2}{R^2} \left( \rho - \bar{\rho} \right),
\label{CLT}
\end{equation}
where $b$ is the impact parameter, $R$ is the areal radius,
\begin{equation}
\rho = 4 \pi \frac{G}{c^2} \frac{ M'}{R^2 R'}, \quad {\rm  and~} \quad
\bar{\rho} = 4 \pi \frac{G}{c^2} \frac{  3M}{R^3} .\nonumber
\end{equation}
We apply the weak field approximations (cf.  Ref. \cite{Bert1966})
and assume that the total Weyl focusing is a sum over all contributors,
$ {\mathcal C} = \sum_i {\mathcal C}_i$. Note that, outside the halo, $\mathcal{C}_i \sim b^2/R^{5}$,
and so only halos that are close to the light ray contribute most significantly.

We again solve Eq. (\ref{dsr}) for $10^6$ different lines of sight.
The Ricci and Weyl focusing are calculated using the halo model, while
$s(z)$ is assumed to be the same as in the FLRW background model [see the first term of (\ref{flrwlimit})].
The resulting distribution for $\Delta$ is presented in Fig. \ref{fig2} (thick solid curve).
 As expected, the variance is much larger than when the smoothed Millennium maps were used. Also, the maximum is shifted more towards positive values of $\Delta$. The mean, as before, is still negligible, at $8\times 10^{-4}$.

For comparison, and to gain a better understanding of the different factors, we now consider rather unphysical variations
of the halo model. The first variation is when we
neglect the shearing -- when solving
 (\ref{dsr}), we put $\sigma =0$. The result of this is shown as the
dashed line in Fig. \ref{fig2}.
As can be seen, it is significantly shifted to the magnification
side (positive $\Delta$) -- the mean is $1.02 \times 10^{-2}$.
This shows that, when considering the halo model,
the role of the shearing should not be neglected when calculating the 
distance correction.

To better understand the 
relation between the halo model and the smoothed Millennium density fields,
we can study two more unconventional modifications of the halo model.
In order to reduce the level of discreteness of the halo model, we 
decrease the value of the parameter $\Delta_{SO}$,
which results in larger halos and a lower 
amplitude of the halo density profiles.
In the first case we reduce $\Delta_{SO}$ to $80$,
which means that the radius of the halo increases by roughly
$30\%$. In the second case, we reduce $\Delta_{SO}$ to $40$,
 which results in a $65\%$ increase in the halo's 
radius\footnote{For example, a halo of mass $10^{12} M_\odot$
and $\Delta_{SO} = 180$
has $R \approx 330$ kpc [this follows from  (\ref{halorad})],
for $\Delta_{SO} = 80$ and $\Delta_{SO} =40$, 
$R \approx 430$ kpc and $R \approx 545$ kpc respectively.}.
We plot these results as dotted lines in Fig. \ref{fig2}.
As can be seen, when we increase the halo radii (and hence decrease
the degree of discreteness), we approach the results obtained
when using the smoothed Millennium maps.

So far, the results reveal the same pattern:
a skewed distribution with a maximum at the demagnification side ($\Delta_{max}>0$).
This can easily be understood within the framework of the weak lensing formula (\ref{DWL}):
when light rays pass through the large-scale structure, they are more likely to propagate through voids than through more dense, compact structures. Therefore, it is more likely to have $\delta <0$
along the line of sight and, as follows from (\ref{DWL}), $\delta < 0 \Rightarrow \Delta>0$.
This is clearly visible when we use the smoothed Millennium maps.
If the smoothing scale is sufficiently large, the resulting density field approaches a Gaussian field
(all non-linearities are washed away, and we recover symmetry between overdense and underdense 
regions), and so the maximum approaches zero (for pure Gaussian fluctuations $\Delta_{max}=0$). In the halo model,
we deal with halos that are very compact and occupy a relatively small volume, which makes propagation through
voids even more likely to occur. We therefore see a shift of $\Delta_{max}$ towards higher demagnification
(see Fig. \ref{fig2}).
Occasionally, light rays pass close to a halo, and then both the Ricci and Weyl focusing
magnify the light bundle. Thus, on average, the total balance is recovered, and the mean of $\Delta$ is almost
zero  (in the lensing framework,
as seen in (\ref{DWL}), due to matter conservation it is exactly zero).
Here we find a mean of order $10^{-4}$.
A small deviation from the expected value in the FLRW model is expected. For example the analogue of the Integrated Sachs-Wolfe effect will
affect the redshift, and thus $(z,D_A)$. 
Also, the local environment plays a role -- note that our
observer is located inside an inhomogeneity (i.e. not within a homogeneous region)
-- see footnote on page 6.

The same pattern has also been reported in previous studies -- see for example in \cite{HoWa1998},
and in ray tracing though N-body simulations.
In fact, our halo model provides very good agreement with the ray tracing through
the Millennium simulation, cf. \cite{MrS1,MrS2}. It is also
in almost perfect agreement with the PDF generated using 
the turboGL code\cite{KKVM2009,KKVM2011}\footnote{The turboGL 
code uses the halo model to calculate the PDF of the weak lensing convergence. The effect of the shear is neglected in turboGL. However, if the zero-shear
PDFs are compared, they are in agreement. Note that, as seen in 
Fig. \ref{fig2}, the zero-shear PDF, apart from a small change in its amplitude, 
looks like the translation of the nonzero-shear PDF.}.

The results are therefore consistent with the most commonly-implemented 
procedures, where the effects of inhomogeneities on light propagation
are calculated using the magnification matrix approach. In almost all cases, one further assumes 
that light propagates on unperturbed geodesics, and so the redshift
(to a source located at some fixed comoving distance) is calculated as in the background homogeneous models,
 while inhomogeneities only affect the magnification/demagnification.
So far, we have been doing the same (that is, we assumed that $s(z)$ is the same as in the background FLRW model).
We therefore examine the effect of changing $s(z)$ in the next section.

\section{The role of the non-uniform expansion rate}\label{sofz}

The last effect that remains to be examined is related to $s(z)$.
So far we have been applying the FLRW formula, but as shown in (\ref{sodz}),
to calculate it correctly we need to solve null geodesics.
We now address this point by studying light propagation 
within the LT Swiss Cheese model.
The advantage of the LT model is  that it is an exact model, and so no 
approximations are required when modelling the light propagation, evolution of perturbations, or the Weyl curvature.
The disadvantage is that each particular inhomogeneity must be spherically symmetric,
and we do not have the freedom to mimic the Millennium density profile. 
Furthermore, we cannot describe virialised objects such as massive matter halos (we will come back to this point later on).
\begin{figure}
\begin{center}
\includegraphics[scale=.9]{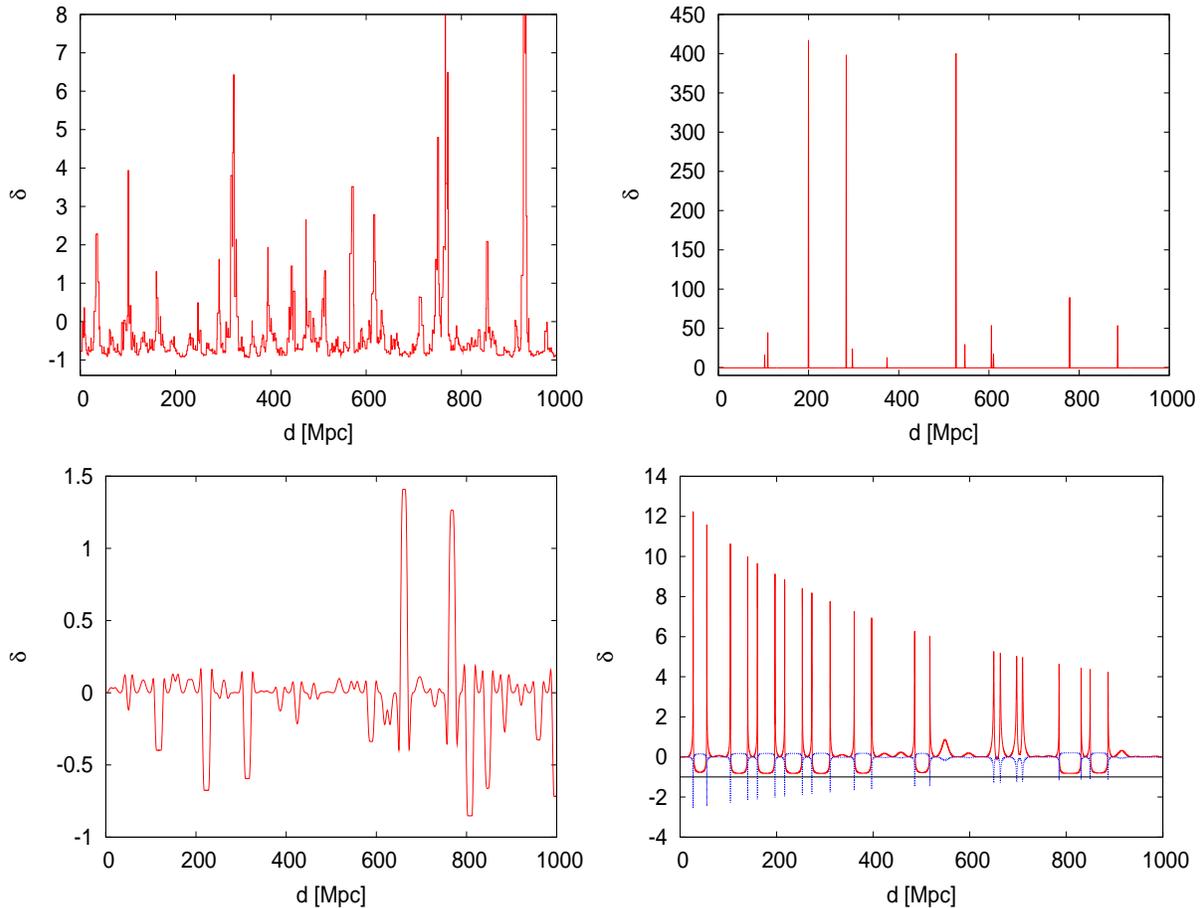}
\caption{{\em Upper left}: Density along a random  line of sight
through the $g_{1.25}$ density field of the Millennium simulation.
{\em Upper right}: 
 Density along a random  line of sight
through the halo model.
{\em Lower left}: 
`Mild' Swiss Cheese model.
{\em Lower right}: 
 Density along a random  line of sight
through the highly non-linear Swiss Cheese model
(solid line -- 'u'-shaped pattern) and the contrast of the expansion rate
$\delta_H = H/\bar{H} - 1$ (dotted line -- note that in regions where $\delta$ is of
high amplitude, $\delta_H$ is negative).}
\label{fig3}
\end{center}
\end{figure}

\begin{figure}
\begin{center}
\includegraphics[scale=1.0]{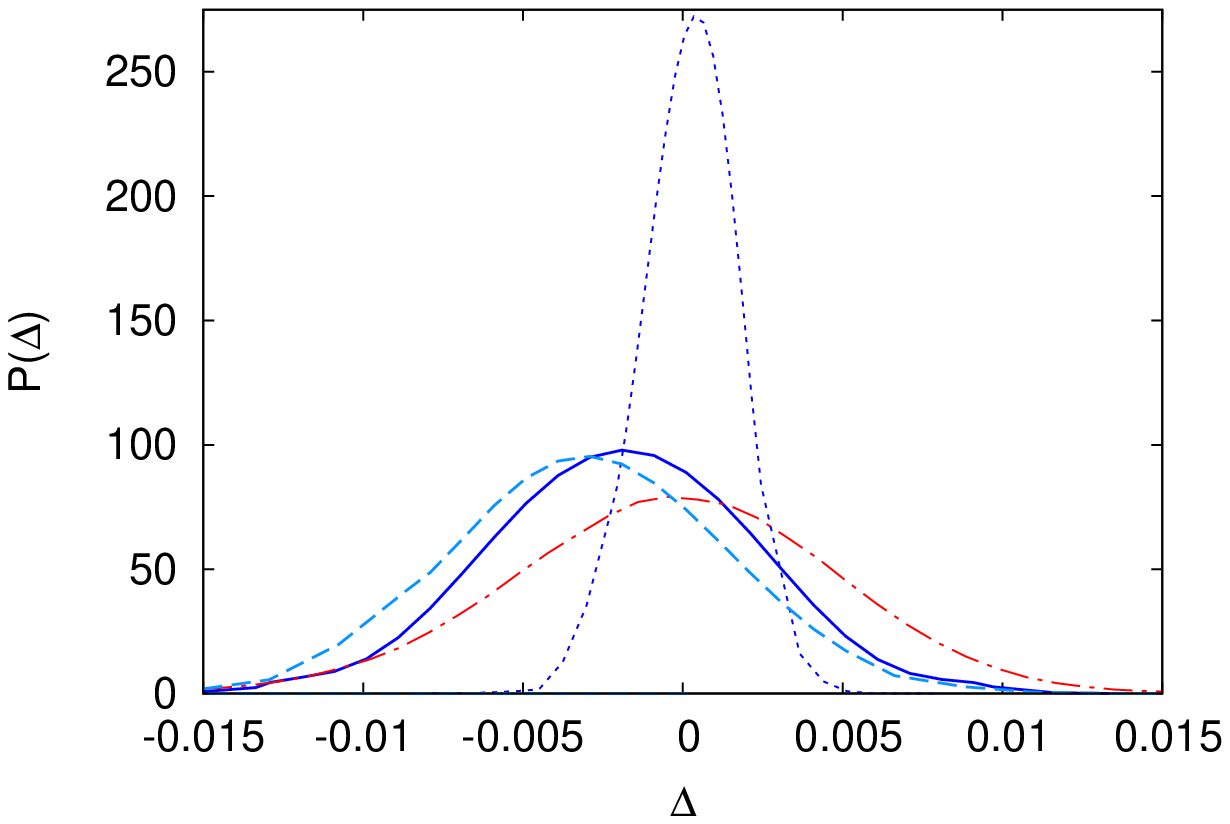}
\caption{PDF of $\Delta$ in various models.
The central dotted line is the PDF obtained within the `mild' Swiss Cheese model.
The PDF of the highly non-linear LT Swiss Cheese model
is given by the solid line. The dashed line is also the PDF obtained
in the non-linear LT Swiss Cheese model, but when we neglect
the effect of collapse on the redshift relation (see text for details).
For comparison, the PDF for the $g_{10}$ density map 
(from Fig. \ref{fig1})
is shown as the dashed-dotted line.}
\label{fig4}
\end{center}
\end{figure}

\setcounter{footnote}{0}

Let us first consider a `mild' Swiss Cheese model,
i.e. one without large density fluctuations. 
Each inhomogeneity has a radius of $20$ Mpc, beyond which the system becomes 
homogeneous.
The density contrast inside the inhomogeneity is generated using the 
log-normal PDF\footnote{$
P(\delta) = \frac{1}{\sqrt{2 \pi \sigma_{nl}^2} } 
\exp \left[ - \frac{ (\ln(1+\delta) + \sigma_{nl}^2/2)^2 }{2
    \sigma_{nl}^2} \right] \frac{1}{1+\delta}$, where
$\sigma_{nl}^2 = \ln [ 1 + \sigma_R^2]$, and
$\sigma_R^2 = \frac{1}{2\pi^2} \int\limits_0^\infty {\rm d} k {\cal P}(k) W^2(kR) k^2$.},
with $\sigma_R = 0.96$ (i.e. $R = 10$ Mpc).
An example of a random line of sight through this model
is presented in the panel to the lower left of Fig. \ref{fig3}.
Within this model, we solve the null geodesic
equations using (\ref{sodz}) to obtain $s(z)$.
The Weyl focusing is calculated from (\ref{CLT}) and the
distance is calculated by solving (\ref{dsr}).
In order to avoid effects of local structures,
we place the observer and the source within the homogeneous
regions. 
The resulting PDF of $\Delta$ for a source at $z=1.6$
is presented in Fig. \ref{fig4} (dotted line).
The mean is 
$ 5.75 \times 10^{-4}$.
Qualitatively, the results are very similar to the ones obtained before:
the PDF is skewed, with the maximum at the positive side of $\Delta$.
Quantitatively, however, the variance is much smaller than in previous
cases.
This is because of the smaller amplitude of inhomogeneities along the line
of sight, which can be seen by comparing
the upper left (Millennium) and upper right (halo model) panels of 
Fig. \ref{fig3}
to the lower left panel of  Fig. \ref{fig3} (the `mild' Swiss Cheese case).

To bring us closer to the Millennium case, but in such a way as to include
shearing and a modified $s(z)$ relation, we now consider a more `realistic' model,
with large deep voids (with, for example, a present day radius of $14$ Mpc and a density contrast of
$-0.83$). These voids have highly non-linear walls which, due to the high density
contrast, are at the collapsing stage, clearly visible in the lower right
panel of Fig. \ref{fig3} (dotted line). 
The PDF obtained within this model is presented in Fig. \ref{fig4} (solid line).
The mean is 
$ -1.01 \times 10^{-3}$.
 The first qualitative difference is that the maximum is now at 
{\it negative} $\Delta$ (magnification).
Secondly, due to large fluctuations in the density field,
the variance is larger, and is comparable
with the $g_{10}$ Millennium density field (dotted-dashed line in  Fig. \ref{fig4}).

\begin{figure}
\begin{center}
\includegraphics[scale=1.]{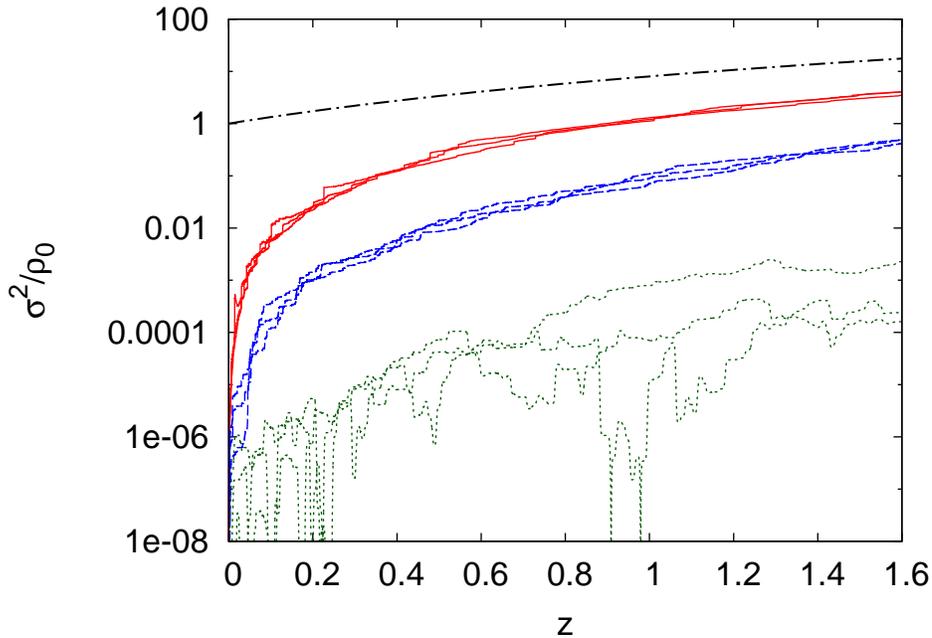}
\caption{ The Weyl focusing (i.e. the ratio of the shearing to present day
matter distribution $\sigma^2/\rho_0$) in the different models that were considered.
From bottom: 
the three dotted lines are 3 examples of the shearing
in the mild LT Swiss Cheese model. The shearing
at $z=1$ is of order $10^{-6}-10^{-4}$.
Above those, the next 3 (dashed) lines 
are 3 examples of shearing in the highly non-linear Swiss Cheese model.
Here, at $z=1$, the shearing is of order $10^{-2}$.
The next 3 (solid) lines show shearing
in 3 examples of the halo model. At $z=1$,
the shearing is of order $10^{-1}$.
For comparison, the uppermost dash-dotted line
is $(1+z)^3$ (the evolution of matter density).}
\label{fig5}
\end{center}
\end{figure}

The surprising result is that the maximum is on the magnification side ($\Delta_{max}<0$).
Such a phenomenon has not been reported before. 
Let us closely examine what could lead to such an unexpected behaviour.
Three working hypotheses can be put forward to explain this: (1) the effect of 
the Weyl focusing, (2) the problem with collapsing regions, (3) an artefact of 
the symmetry of the density contrast.
With regards to the shearing, we have seen in the previous section,
when comparing the halo model with and without the Weyl focusing,
that if $\sigma\ne0$, the PDF shifts to lower values of $\Delta$.
However, as seen from Fig. \ref{fig5}, the role of the Weyl focusing
for our Swiss Cheese models is small 
(cf. \cite{Szyb2011} and \cite{FKWV2012} for a similar conclusion
regarding the role of the Weyl focusing in this type of model).
Essentially, if the density contrast is not high (unlike
in the halo model; compare the upper right panel of Fig. \ref{fig3} with the 
lower right panel),
the role of the Weyl focusing is small (compare the set of 3 solid lines
with the set of 3 dashed or dotted lines in Fig. \ref{fig5}).
We therefore conclude that the shear cannot be responsible
for the shift of $\Delta_{max}$ from demagnification to the magnification side.
This can be confirmed by re-running the same model
with $\sigma$ set to zero -- the resulting PDF is exactly the same as 
in the case of $\sigma\ne0$ (solid line in Fig. \ref{fig4}).

Our second hypothesis involves considering the role of collapsing regions.
In this case, $H<0$, which may strongly affect $s(z)$.
In the real Universe, whenever the density field has a high
amplitude, cosmic structures are virialised, and hence
they are not collapsing. Unfortunately, there is no rotation within the LT 
model, which could prevent the collapse.
Therefore, as seen in the lower right panel of Fig. \ref{fig3}, the
expansion is negative within the wall.
In order to `correct' the model for the virialisation effect,
whenever the expansion rate is negative, we assume that the structure is
already virialised and so the redshift does not change within this region.
That is, we assume that $ {\rm d} k^0 / {\rm d} s = 0$ in these regions,
whereas in all other regions we use the exact formula,
i.e. $ {\rm d} k^0 / {\rm d} s = - R \dot{R} (k^3)^2 - R'
(\dot{R}'/(1+2E)) (k^1)^2 $.
We should exercise caution
when analysing these results, as this `correction' was done by hand.
The fact that some regions are  virialised, and hence $ {\rm d} k^0 / {\rm d} s = 0$, is not
a consequence of a model, and so there is clearly an issue of self-consistency.
The resulting PDF is presented using a dashed line in Fig. \ref{fig4},
where we find that the maximum of the curve is shifted even further towards negative values of $\Delta$.
The mean is almost twice as large as before,  
 $ -1.91 \times 10^{-3}$.
Clearly this does not explain why $\Delta_{max} <0$, and in fact it makes
things even `worse'.

Our final hypothesis concerns the symmetry of the density 
 perturbations. We can 
already see that this cannot
be the cause, as $\Delta_{max} >0$ for the `mild' Swiss Cheese model.
Nevertheless, we return to the Millennium smoothed maps and perform further tests.
We proceed as before, with the only difference 
being that, when solving $s(z)$, we no longer assume a uniform expansion rate 
as in (\ref{flrwlimit}), but instead assume that 
\begin{equation}\label{sofzdh}    
 \frac{{\rm d} z}{{\rm d} s} = (1+z)^2 \bar{H} (1+\delta_H), 
\end{equation}
where $\bar{H}$ is the expansion rate of the background model.
In general 

\[\bar{H} \delta_H = \frac{1}{3} \left({\Theta- \bar{\Theta}}\right)
+ \sigma_{ab} e^a e^b \]
where $\Theta$ is the matter scalar of the expansion
and $\sigma_{ab}$ is the shear of the matter velocity field. 
We further work under the assumption of zero shear
and so we only consider fluctuations in the expansion field,
which in the linear regime are given by
\begin{equation}
\delta_H = -\frac{1}{3} f \delta.\label{deltaH}
\end{equation}
Here, $f$ is the growth function, $f = \dot{{\cal G}}/({\cal G}H)$.
This method has thus 2 limitations,
and works well as long as (1) the density contrast
is not too high and (2) the matter shear is negligible.
We proceed further with this toy model, as it will
bring more insight into results obtained within the exact LT model.

\begin{figure}
\begin{center}
\includegraphics[scale=0.98]{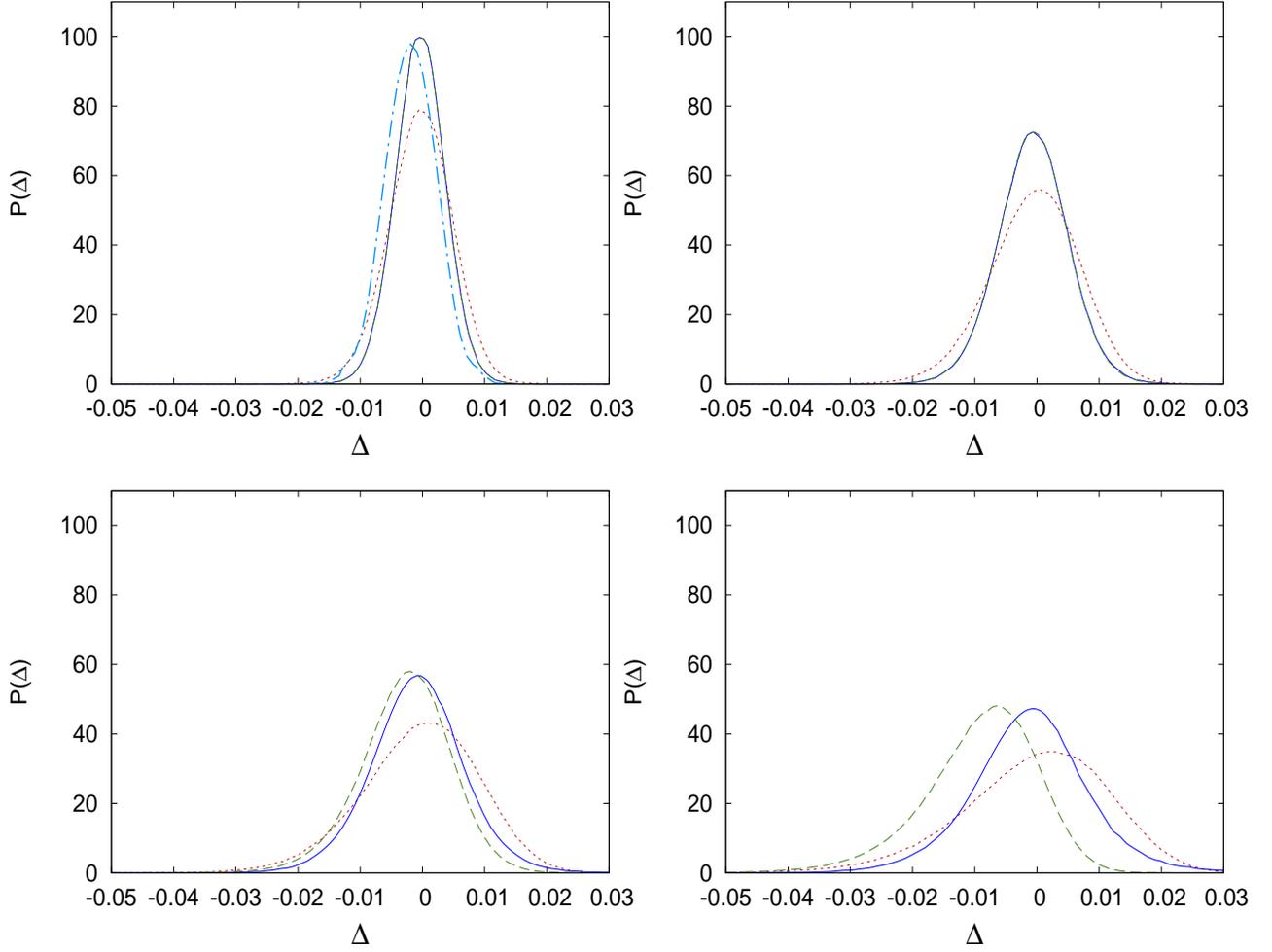}
\caption{
{\em Upper left}: Millennium map $g_{10}$ with $\delta_H =0$
(dotted line, cf. Fig. \ref{fig1}),
$\delta_H \ne 0$ (solid line), 
and $\delta_H \ge -1$ (dashed line, in this case indistinguishable from the solid line).
For comparison, the PDF of the non-linear Swiss Cheese model is shown
as the dash-dotted line.
{\em Upper right}:
Millennium map $g_{5}$ with $\delta_H =0$
(dotted  line),
$\delta_H \ne 0$ (solid line),
and $\delta_H \ge -1$ (dashed line, hardly distinguishable from the solid line).
{\em Lower left}: 
Millennium map $g_{2.5}$ with $\delta_H =0$
(dotted  line),
$\delta_H \ne 0$ (solid line),
and $\delta_H \ge -1$ (dashed line).
{\em Lower right}: Millennium map $g_{1.25}$ with $\delta_H =0$
(dotted  line), 
$\delta_H \ne 0$ (solid line),
and $\delta_H \ge -1$ (dashed line).}
\label{fig6}
\end{center}
\end{figure}

Naively, one can think of this as follows: because $\av{\delta} = 0 = \av{\delta_H}$,
we should expect (qualitatively) the same results as before, except
for a potentially larger variance due to the presence 
of both $\delta$ and $\delta_H$.
Our results are presented in  Fig. \ref{fig6}.
First of all, the variance changes only slightly.
Secondly, and most importantly, we find that $\Delta_{max}<0$.
In the $g_{10}$ case (with 10 Mpc smoothing scale), the results
are strongly comparable with our non-linear Swiss Cheese model; 
for comparison we present its PDF in the upper left panel of Fig. \ref{fig6} (dot-dashed line). 
Also, the mean of $\Delta$ is of the same order of magnitude as in the
Swiss Cheese model, i.e. $\sim 10^{-3}$ --- see Table \ref{tab1}.

Finally, as before, we try to `correct' the model for the effects of virialization.
As seen from (\ref{deltaH}), if
\[ \delta > \delta_T = \frac{3}{f}, \]
then such a region collapses. In the $g_{1.25}$ case, almost $1.5\%$
of regions have a density contrast above this threshold (see Table \ref{tab2}).
As before, we set 
\[ {\rm if~} \delta > \delta_T \quad \Rightarrow \quad \delta_H = -1.\]
In this case, if we calculate the average $\delta_H$,
due to the fact that we neglect the most negative contributions, the average is 
no longer zero, and in the case of $g_{1.25}$ 
the average $\delta_H$ is approximately $0.0211$ (it can only be zero
if $\delta_H < -1$).
The resulting PDFs are presented in Fig. \ref{fig6} (dashed lines).
As in the case of the non-linear Swiss Cheese model before,
the PDF shifts towards even more negative
values of $\Delta$. For detailed values of the means of the PDFs, see Table \ref{tab1}.

\begin{table}
 \caption{The mean and standard deviation of $\Delta$
for a source at $z=1.6$.}
\begin{center}
\begin{tabular}{cccc}
  Model & $\delta_H =0$  & $\delta_H \ne 0$ &  $\delta_H \geq -1$ \\ \hline
 $g_{1.25}$ & $1.20 \times 10^{-4}$ &  $-4.71 \times 10^{-4}$ & $-8.77 \times 10^{-3}$\\
 $g_{2.5}$ & $1.38 \times 10^{-4}$  &  $-2.85 \times 10^{-4}$ & $-2.44 \times 10^{-3}$\\
 $g_{5}$ & $1.29 \times 10^{-4}$  & $-1.43 \times 10^{-4}$ & $-2.57 \times 10^{-4}$\\
 $g_{10}$ & $1.44 \times 10^{-4}$  & $-2.18 \times 10^{-5}$ & $-2.62 \times 10^{-5}$\\
mild Swiss Cheese & $5.75 \times 10^{-4}$  & $-$ & $-$\\ 
non-linear Swiss Cheese & $-$  & $-1.01 \times 10^{-3}$ & $-1.91 \times 10^{-3}$\\
 halo & $-8.02 \times 10^{-4}$  & $-$ & $-2.31 \times 10^{-2}$
\label{tab1}
\end{tabular}
\end{center}
\end{table}

\begin{table}
 \caption{
{\em Middle column}: The percentage of regions within the Millennium simulation
whose density contrast is higher than the threshold above which $\delta_H < -1$
{\em Right column}: The average expansion rate after 
setting $\delta_H = -1$ wherever $\delta_H<-1 $.}
\begin{center}
\begin{tabular}{ccc}
  Map & $\%$ of $\delta> \delta_T$ & $\av{\delta_H}_T $\\ \hline
 $g_{1.25}$ & $1.459 \%$ &  $1.95 \times 10^{-2}$ \\
 $g_{2.5}$ & $ 0.762 \%$  &  $5.31 \times 10^{-3}$ \\
 $g_{5}$ & $ 0.089\%$  & $2.69 \times 10^{-4}$ \\
 $g_{10}$ & $0$  & $0$ \\
\label{tab2}
\end{tabular}
\end{center}
\end{table}

\begin{figure}
\begin{center}
\includegraphics[scale=0.55]{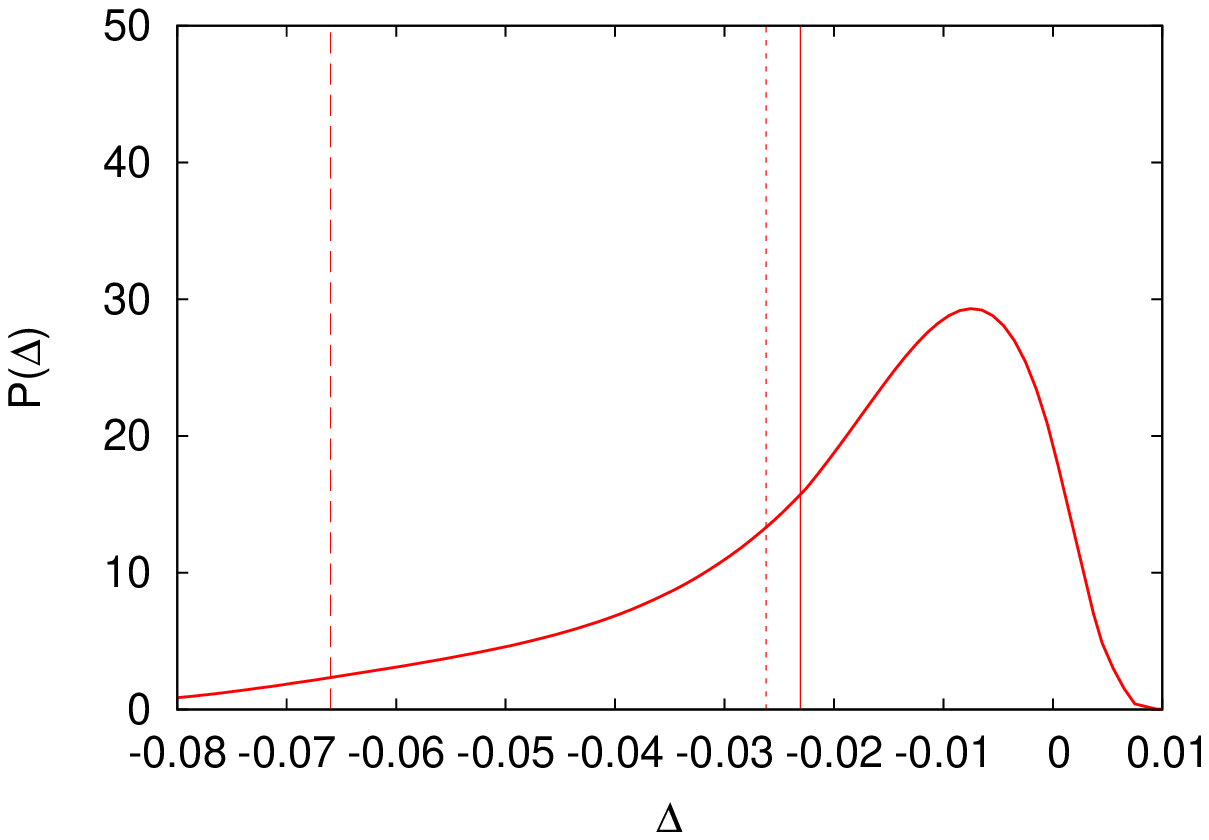}
\includegraphics[scale=0.55]{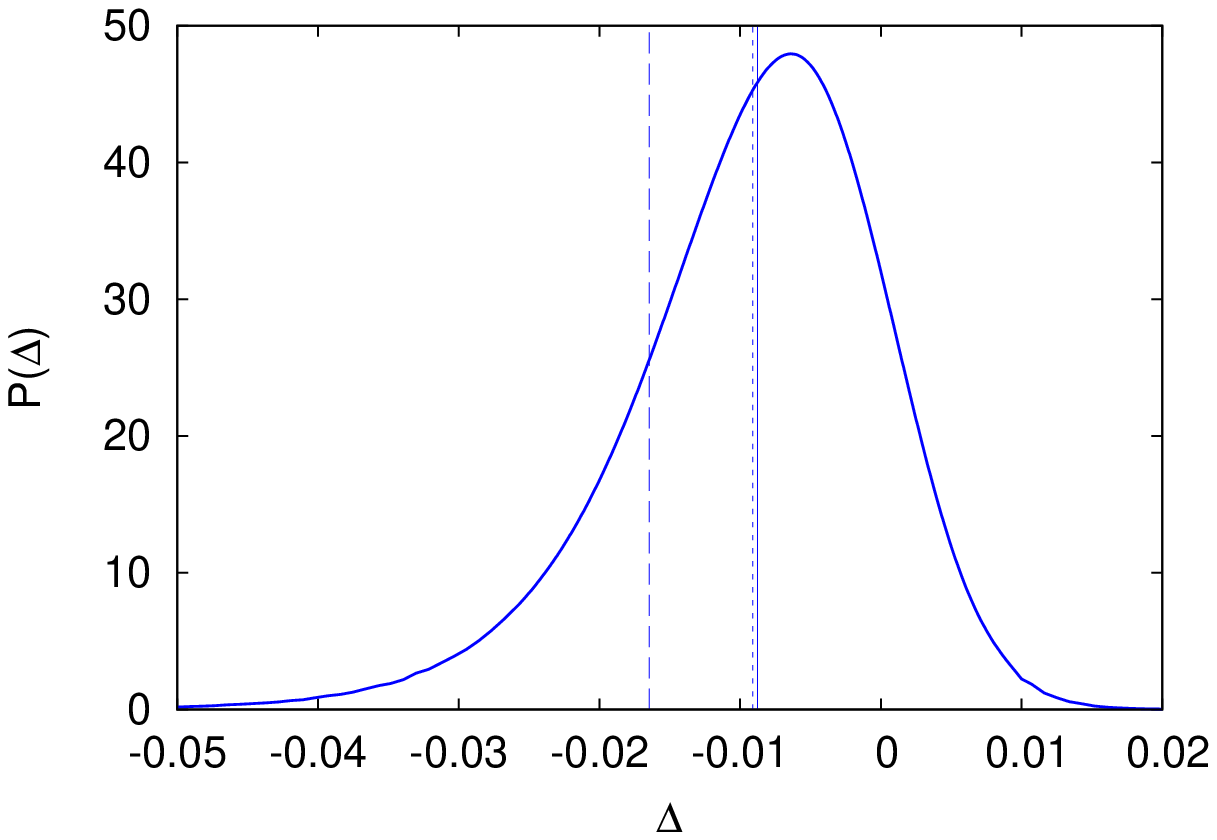}
\caption{PDF of $\Delta$.
Solid line: mean. Dotted line: solution of (\ref{approxsetD}).
Dashed line: when (\ref{flrwevo}) is used together with the assumption
that $\tilde{z} = z$. {\em Left}: halo model, {\em Right}: $g_{1.25}$ Millennium map.}
\label{fig7}
\end{center}
\end{figure}

Proceeding further with the idea of a variable expansion rate and its
impact on $s(z)$, we also recalculate our results for the halo models that were 
considered previously.
Since halos are virialised, their interiors are set to $\delta_H = -1$.
The resulting PDF is presented in the right panel of Fig. \ref{fig7}.
As was the case for the Millennium maps, a shift towards magnification is evident.
This should be compared with the mild Swiss Cheese model, or with the
models which assumed a uniform expansion rate, where the
PDF peaked on the positive side of  $\Delta$.

\begin{figure}
\begin{center}
\includegraphics[scale=0.7]{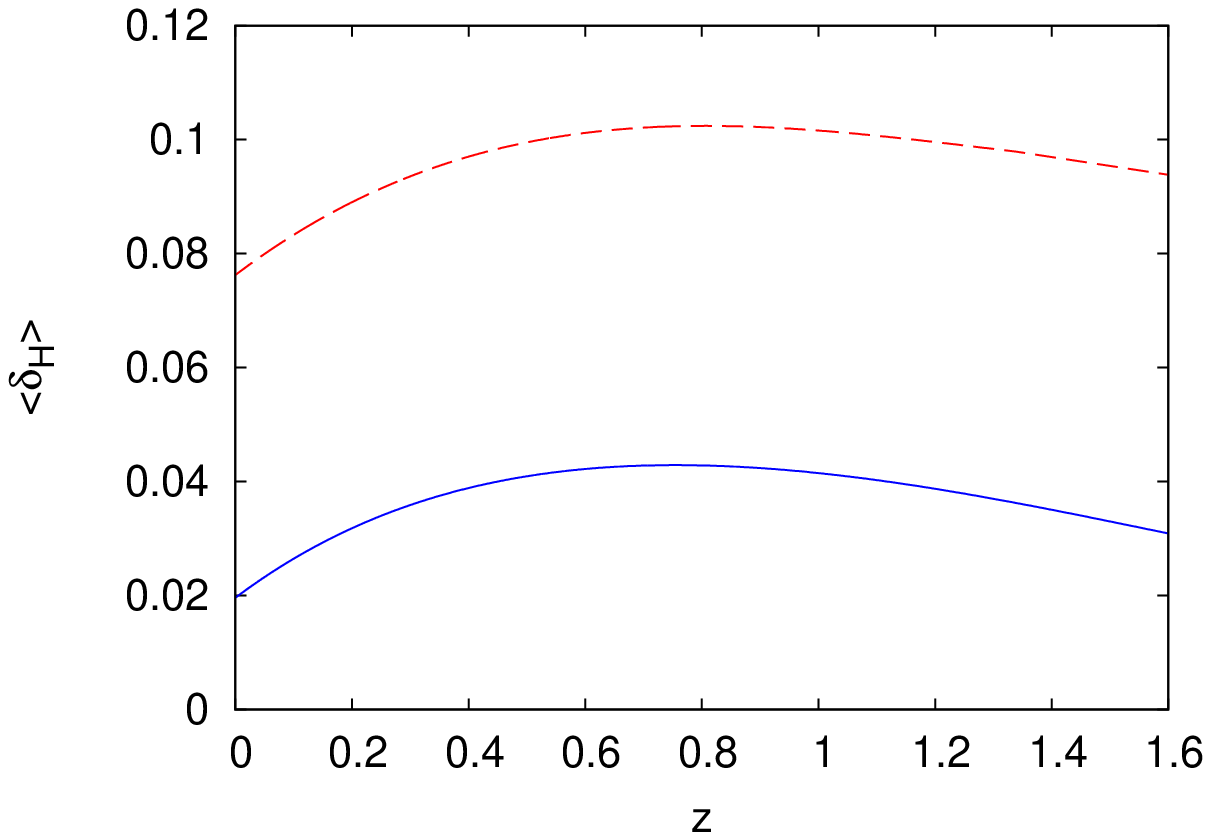}
\caption{Average of the fluctuation of the expansion rate 
at constant $z$. Solid line 
$g_{1.25}$ Millennium map, dashed line for the halo model.}
\label{fig8}
\end{center}
\end{figure}

In Secs. \ref{R} and \ref{wsigma} we 
assumed that $s(z)$ was the same as in the FLRW background
model. As a result, we found that $\av{\Delta} \approx 10^{-4}$.
For the Swiss Cheese model, we had variations in $s(z)$
with regards to the background, but still found that $\av{\Delta} \approx 10^{-3}$.
It was only when we started to `correct' our models for virialization that
we found deviations of order of $10^{-2}$ (for the halo and $g_{1.25}$ models).
We now want to follow this up further, and test how the
average of the distance follows the background's $s(z)$.
Note that when we `force' some regions not to collapse (by setting $\delta_H = -1$ whenever $\delta_H < -1$), we change the average expansion history 
(see Table \ref{tab2} and Fig. \ref{fig8}),
and thus also $s(z)$.
This leads us to question: if we recover the average $s(z)$ correctly, will we 
obtain the correct 
mean of the distance, without the need for a 
detailed calculation of the Ricci and Weyl focusing?

We study this issue within the halo and $g_{1.25}$ models,
as they have the largest deviation of $\av{\Delta}$.
The PDFs obtained within these models are presented 
in Fig. \ref{fig7}. The mean of $\Delta$ is shown
as a vertical solid line.
The average of the expansion rate fluctuations calculated on a slice of constant $z$
is presented in Fig. \ref{fig8}.
 Before proceeding further, we must comment on one important aspect
of our computations. Cosmologists tend to use the redshift as a proxy for time,
so they express the evolution of different fields in terms of the redshift, for example:
\begin{eqnarray}
 \rho(s) \Leftrightarrow  \rho(\tilde{z}) = \rho_0 (1+\tilde{z})^3, \nonumber \\
 H(s) \Leftrightarrow H(\tilde{z}) = H_0 \sqrt{\Omega_m (1+\tilde{z})^3 + \Omega_k  (1+\tilde{z})^2 + \Omega_\Lambda}.
\label{flrwevo}
\end{eqnarray}
Only in the case of perfect homogeneity do
we have that $\tilde{z}$ 
(the proxy for time, i.e. the redshift of a photon at a given time in a perfect homogeneous model) 
coincides with $z$ (the actual shift of the photon's frequency). 
Thus, in solving the Sachs equations, one must solve the following:
\begin{eqnarray}
&& \frac{{\rm d^2} D_A}{{\rm d} s^2} = -  \frac{1}{2} \av{\rho} (1+z)^2 =  \rho_0 (1+\tilde{z})^3 (1+z)^2 \nonumber \\
&&   \frac{{\rm d} z}{{\rm d} s} = (1+z)^2 \av{H} = (1+z)^2  H(\tilde{z}) (1+\delta_H(z)) \nonumber \\
&&   \frac{{\rm d} \tilde{z}}{{\rm d} s} = (1+\tilde{z})^2 H(\tilde{z}).
\label{approxsetD}
\end{eqnarray}
We solve these, and plot the results as dotted lines in Fig. \ref{fig7}.
As can be seen, the difference between the actual value, and that obtained
from the above set of equations, is of order of $10^{-3}$,
(as before, where we set $s(z)$ to its value in the background $\Lambda$CDM model).
For comparison, we solve the above equation with 
\[ z = \tilde{z} \quad \Rightarrow \quad R_{\alpha \beta} k^\alpha k^\beta 
= \rho_0 (1+z)^5. \]
This time, the result (given by the dashed line) is significantly displaced 
to the left. The reason for the shift towards higher magnification 
is that if $\av{\delta_H} > 0$ then, as follows from (\ref{approxsetD}), we have that
$z > \tilde{z}$, and so
if we replace $\tilde{z}$ with $z$, we increase the Ricci focusing, hence
higher magnification.
This simple exercise shows that if one is interested in having an average
$D_A(z)$ without having to worry about the Ricci or Weyl focusing,
one can simply use the average $s(z)$ of one's model.
Caution must be taken, though, when studying models
in which 
$\rho$ is not proportional to $(1+z)^3$.
Such models include, for example,  
Gpc-scale inhomogeneous LT models\footnote{In fact the results
of this paper provide a fresh view into different configurations of the LT
model, which consist of giant voids \cite{BKHC2009,BCK2012}
and giant humps \cite{CBK2010}. 
Within the giant voids $\rho$ increases outwards,
which results in the decrease of the expansion rate along the past null cone,
hence the dimming of the supernova within the giant voids
is caused by the change of the $s(z)$ relation. 
Whereas for the giant hump models
 $\rho$ decreases outwards making the Ricci focusing less efficient,
and hence causing additional dimming.}
 and some backreaction models.

In this section we have tested all 3 factors that affect the relation between $z$ and $D_A$
via the Sachs equation: $\rho(z)$, $\sigma(z)$, and $s(z)$.
In order to model them in a self-consistent manner, we used the LT Swiss Cheese model.
In all cases, we found that the distance does not change, on average, by more 
than $10^{-3}$ (at $z = 1.6$).
This result is fairly consistent with what has been found 
by other groups who have employed exact models,
and ensured the proper randomization of light rays\footnote{Note that 
for models where the distance correction is calculated using the Born approximation, have by the construction $\av{\Delta}=0$ \cite{VaFW2008,KKVM2009,KKVM2011,FKWV2012}}
(cf. \cite{BrTT2007,BrTT2008,ClZu2009,Szyb2011,MNBM2012,DVD2012}).

Our main motivation has been to follow, as closely as possible, the standard procedures
to test self-consistency and the impact of inhomogeneities on the optical properties of the Universe within models that follow the FLRW evolution. Nevertheless, it is not completely obvious that 
one should recover the FLRW results for these models,
unless one completely agrees with Weinberg's reasoning.
If one {\it does} take Weinberg's results at face value, then 
our $10^{-3}$ results should be considered as a {\em big}
deviation, about two orders of magnitude larger than what is allowed
by the analogue of the Integrated Sachs Wolfe effect (which should be of order $10^{-5}$).

From one point of view, our results have turned out to be quite surprising.
In the case of large fluctuations in the density and expansion 
fields, the mode of $\Delta$ shifts to the magnification side.
Such a behaviour has not been reported before.
The effect is only visible if the degree of inhomogeneity is large --
for small amplitude Swiss Cheese models, we recover the standard results.
This can be explained using the same `voids argument' that we used
in the case of the weak lensing approximation --
since the structures in the Universe form a cosmic web, it is more likely
for photons to propagate through voids (hence the position of the mode is associated with the amount of void regions).
As seen from (\ref{DWL}), the weak lensing regime is only sensitive to density fluctuations, and so the mode is positive. If we introduce fluctuations in the expansion rate, then
the redshift decreases, as seen from (\ref{sofzdh}).
Hence an object located at the same distance has a higher redshift,
and thus gives the impression of being magnified compared
to the homogeneous case\footnote{The same fact can also be visualised
in terms of the slope of the relation $(z,D_A)$ -- when
the redshift increases faster than in the homogeneous case, the slope decreases. See, for example, Fig. 4 of \cite{MNBM2012}, which shows
the change of the slope in the relation $(z,D_A)$,
depending on fluctuations in the density and expansion rate.}.
This is the reason why the mode is now on the magnification side.

A valid point, and one worth further investigation, is related to the fact 
that, due to the virialisation of high density
regions, the average expansion rate may deviate from the background value.
This was already pointed out in \cite{RaSy2006,RaSy2008}
in the context of backreaction models,
although as shown in \cite{MM2010,MM2011}, the presence
of matter shear may decrease the backreaction effect,
leaving the evolution of the average quantities relatively unchanged.
Nevertheless, in such cases, the 3D volume average
on surfaces of constant time may not coincide  
with the average on slices of constant $z$.
As a result, as we have shown above, we would still expect a change in the redshift-distance relation.
However, we reiterate that care should be taken in interpreting these results -- 
changing $H$ by hand may
not lead to self consistent results (although keeping the expansion rate
uniform may also turn out to be inconsistent).
To test this phenomenon more closely, we require exact models that allow for 
virialisation. The LT models are definitely not suitable for such a study.

\section{Conclusions}
\label{disc}
The overarching question that we wish to address is: does the FLRW model, which correctly describes the evolution of the background, 
correctly recover (on average) the distance-redshift relation? Our results, while not entirely conclusive, are suggestive. We have explored a suite of models which are inhomogeneous, but close to the FLRW. We find that, if the background model correctly describes the evolution of the average expansion rate and the average matter density, then indeed the homogeneous model is a good approximation.

Our results show that  Ricci focusing alone introduces deviations from the average value. The PDF is skewed and the maximum is on the demagnification side
(a random object will most likely be dimmer than the average).
The change of the mean is negligible,
which means that, on average, distances coincide with the expected distance
in the background model.
The deviation from the average (for a particular line of sight)
depends crucially on the smoothing scale and, unsurprisingly, the smaller the smoothing scale, the larger the deviation. We find that Weyl focusing is important for objects with very high density contrast, and yet even in this case, the bias in the mean of $\Delta$ is negligibly small.

We also find that the presence of large fluctuations in the expansion rate also affects the final results:
 there is a slight shift of the maximum of the PDF towards
the magnification side (a random object will most likely be brighter than the average), and the variance changes. In particular, we find that if the average expansion rate deviates from the background expansion rate, the distance (on average) changes by a few percent.
In the real Universe, such a situation may occur in the late non-linear stage of evolution, where large voids expand much faster than the background, while virialised over-dense regions do not expand at all 
(for these fluctuations to cancel each other, at some stage the presence of collapsing regions is required).

In summary, we have studied the various biases that cosmological inhomogeneities can introduce into the inferred optical properties of the Universe by considering a range of toy models. There are, of course, limitations to the accuracy and self-consistency of these models, and how well they mimic the real Universe. We find that the presence of inhomogeneities (around a fixed background)
changes distances by at most a few percent, although the average is closely related 
to the background model. The only possible deviation of the average of the distance
from the distance predicted by the background model
is in situations where the background model does not correctly describe the evolution of the average quantities.

 The lessons we have learned can be summarized in the form of a message to our readers. 
 For cosmologists, we would like to point out that, with the increasing precision of cosmological observations, a 
change of few percent might soon become important for accurate 
estimation of cosmological parameters (cf. \cite{LKMQ2010,kb2011c}).
In particular, the PDFs are skewed and hence, for a limited 
number of observational data, there might be a bias due to the 
fact that the mode does not coincide with the mean, or median.
For relativists, we would argue that, to study the effect of inhomogeneities on observations, it is first
important to understand how much
inhomogeneities affect the evolution of the background model.
Such a study should focus on $\rho(z)$ and $H(z)$, and
especially on potential deviations from (\ref{flrwevo}),
as this has the biggest impact on the distance-redshift relation
(see, for example, Fig. \ref{fig7} for a few percent change
in the case where $\rho(z) \nsim (1+z)^3$ (cf. \cite{SyRa2011}).
Thus, pinning down $\rho(z)$ and $H(z)$ will also provide
the average of $D_A(z)$, without the need to account for the Weyl focusing
or to model complex aspects of the Ricci focusing).

\ack

We acknowledge discussion with Timothy Clifton, 
Sabino Mataresse, and Chris Clarkson. We are particularly grateful to
Phil Bull for a detailed reading of the manuscript and extensive discussions.
This research of was supported by 
the European Union Seventh Framework Programme under
the Marie Curie Fellowship, grant no. PIEF-GA-2009-252950, STFC, BIPAC, and the
Oxford Martin School.

\end{document}